%% file: report.tex
\documentclass[letterpaper,11pt]{report}

\pdfoutput=1

\title{A Java Data Security Framework (JDSF) for MARF and HSQLDB}

{\author
	{
		Serguei A. Mokhov\\
		Lee Wei ``Lewis'' Huynh\\
		Jian ``James'' Li\\
		Farid Rassai\\\\\\\\\\
		Concordia University\\
		Montr\'eal, Qu\'ebec, Canada\\\\\\
	}
}

\date{\input{date}}

\usepackage{graphicx}
\usepackage{latexsym}
\usepackage{makeidx}
\usepackage{url}
\usepackage{hyperref}

\makeindex

\input{styles}

\newcommand{\dmarf}[0]{DMARF\index{MARF!Distributed}\index{Frameworks!Distributed MARF}\index{Libraries!Distributed MARF}}
\newcommand{\jdsf}[0]{JDSF\index{Frameworks!JDSF}\index{Libraries!JDSF}}
\newcommand{\hsqldb}[0]{HSQLDB\index{HSQLDB}\index{Tools!HSQLDB}\index{Databases!HSQLDB}}

\begin{document}

	\begin{titlepage}
		\maketitle
	\end{titlepage}

	\include{toc}
	\include{introduction}
	\include{methodology}
	\include{conclusion}
	\include{references}

	\printindex
\end{document}

%% file: date.tex
May 2007

%% file: styles.tex
\newcommand{\marf}{\textsf{MARF}}

\newcommand{\xf}[1]{Figure~\ref{#1}}

\newcommand{\xs}[1]{Section~\ref{#1}}

\newcommand{\gipsy}{{GIPSY\index{GIPSY}}}

\newcommand{\C}{{C\index{C}}}

\newcommand{\file}[1]{\texttt{#1}\index{Files!#1}}
\newcommand{\tool}[1]{\texttt{#1}\index{Tools!#1}}
\newcommand{\api}[1]{\texttt{#1}\index{API!#1}}

%% file: toc.tex
\pagenumbering{roman}
\tableofcontents
\clearpage
\pagenumbering{arabic}

\listoffigures

%% file: introduction.tex
\chapter{Introduction}
\index{Introduction}

$Revision: 1.1.2.10 $

This project explores secure data storage related issues from the point of
view of database security in two open-source projects: {\marf}~\cite{marf} and HSQLDB~\cite{hsqldb} and
proposes a relatively independent reusable framework to enable database security
features in both projects. While we present a comprehensive list of the
things that can be done in this project through the related work and literature
review, we cover a
subset of it, iteratively~\cite{jdsf-integrity-cisse08,%
jdsf-authentication-cisse08,%
jdsf-api-poster,%
sqlrand-key-management}.

\section{{\marf}}

The following quoted from
\url{http://marf.sf.net}, \cite{marf}, as a brief introductory note.

\begin{quote}
{\it
Modular Audio Recognition Framework (MARF) is an open-source research
platform and a collection of voice/sound/speech/text and natural language
processing (NLP) algorithms written in Java and arranged into a modular
and extensible framework facilitating addition of new algorithms. MARF can
run distributively over the network (using CORBA, XML-RPC, or Java RMI) and
may act as a library in applications or be used as a source for learning
and extension. A few example applications are provided to show how to use
the framework. One of MARF's applications, \cite{SpeakerIdentApp}~\cite{marf-speaker-ident-app} has a database
of speakers, where it can identify who people are regardless what they
say.}~\cite{marf}
\end{quote}

\noindent
Original {\marf} was designed and developed by
Mokhov et al. and
collaborators
throughout a variety of projects~\cite{marf,marf02,dmarf06}.

\clearpage

\section{HSQLDB}

The following quoted from
\url{http://hsqldb.org/}:

\begin{quote}
{\it
HSQLDB is the leading SQL relational database engine written in Java. It
has a JDBC driver and supports a rich subset of ANSI-92 SQL (BNF tree
format) plus SQL 99 and 2003 enhancements. It offers a small (less than
100k in one version for applets), fast database engine which offers both
in-memory and disk-based tables and supports embedded and server modes.
Additionally, it includes tools such as a minimal web server, in-memory
query and management tools (can be run as applets) and a number of
demonstration examples.

The product is currently being used as a database and persistence engine
in many Open Source Software projects and even in commercial projects and
products. In it's current version it is extremely stable and reliable. It
is best known for its small size, ability to execute completely in memory,
its flexibility and speed.

This feature-packed software is completely free to use and distribute
under our licenses , based on the standard BSD license. Completely free of
cost or restrictions and fully compatible with all major open source
licenses. Java source code and extensive documentation included.}~\cite{hsqldb}
\end{quote}

\section{Long Term Goals}

We propose to provide a framework to ensure the data privacy, integrity,
and authentication aspects for {\marf}'s and HSQLDB database(s). {\marf} would be the
front-end (FE) of HSQLDB. Depending on the architecture, {\marf} can be a trusted
or untrusted FE and so is the HSQLDB instance. When HSQLDB does not trust
{\marf} is on the level of SQL injection coming from the {\marf}'s data.
A framework approach is required as to be able to compare and select
better algorithm implementations from the available pool of implementations.

There are several ways to do it, several architectures, algorithms, etc.,
so on the research side of the project, we research on
several techniques to achieve the required goals, compare them, and provide a
framework implementation-wise such that it is easy to add new algorithms that
implement the goals, but for the project only implement one to three) of
those techniques within the designed framework as a proof-of-concept.

To summarize:

\begin{itemize}
\item
we read a few research papers on the techniques for
 privacy/integrity/authentication of the data storage, and schema
 randomization.
\item
propose and design a framework that allow easy plugging-in of
 such implementations within {\marf} and HSQLDB, with the API, etc.
\item
implement a few of such techniques and briefly compare them for the results
 (e.g. complexity, performance, strength, trust model, etc.)
\item
there is an emphasis on doing this for a high volume of a multimedia data
 (in this case for now mostly audio and text)
\end{itemize}

\subsection{Proposed Implementation Details}

MARF is using a some sort of database to store speaker identities and a
mapping to their voice samples. Regardless the storage model (relational
(MARF can use a connection to any relational database, e.g. PostgreSQL,
HSQLDB, MySQL, etc.) through an appropriate JDBC driver) or plain Java
objects (default), or XML), the data travels between the library
components and the applications to storage a ``plain text''.

One of the applications of MARF collects various performance statistics of
the implemented algorithms in the framework, and ranks them against each
other. This type of statistics does not require the tester or scientists
to know the exact identities of speakers and link them to the specific
voice samples. Since we are talking about identities of speakers and we
collect statistics in MARF, how can we conceal the identities, while
providing meaningful non-disclosing stats?

We may also do not trust the underlying storage model to provide privacy
of the MARF's databases, we propose to implement a layer at the MARF's
library layer to provide some privacy, integrity, and authenticity checks
through any available optional cryptographic framework so users interested only in
statistics or optimization tuning cannot snoop on the actual
data-in-transit, and yet, get useful results for their research or
otherwise applications.

If we pick HSQLDB as the backend database engine for MARF and its
applications, we can make it trusted as both components and the new
framework are our works. While HSQLDB has a comprehensive implementation
of features, there is a room for improvement to tighten security in
HSQLDB: (1) implement privacy (encryption, maybe using one of the
k-anonymity techniques) per column-basis to prevent DBAs snooping on the
sensitive data, (2) implement schema randomization to
enhance the defense against SQL injections (e.g. we do not trust MARF's
data). Thus, we propose to implement at least one of these aspects as time
permits, and release our contribution back to the open-source community of
HSQLDB.

\subsection{Assumptions}

A this stage, this projects is exclusively looking into the confidentiality (privacy),
integrity, and authentication of the {\em data} or its origin in some form of a database.
There are no users or clearance levels in this model, so there are no
issues of authorization and access control, multilevel databases. Nor
we consider the availability aspect.

\subsection{Resources}

While implementing, integrating, and testing framework, we may borrow
open-source implementations of known crypto and otherwise algorithms,
with due credit
given to the original developers. We will have to do the work of adapting
those algorithms into our structure.

As a part of the process, we:

\begin{enumerate}
\item
Read and summarize the relevant research papers.

\item
Framework design, including base interfaces and data structures.

	\begin{enumerate}
	\item
  Privacy aspect
	\item
  Integrity aspect
	\item
  Authentication aspect
	\item
  SQL/schema randomization aspect
	\end{enumerate}

%

\end{enumerate}

\section{Literature Review}

This section presents the summary of the research done, such
that the framework being developed covers most aspects and parameters to
be flexible. We will review the building blocks of any security-aware
information system, such as CIA$^N$ -- primarily confidentiality,
integrity, and authentication of {\em data}. We will not address availability,
authorization, and access control aspects in this work.

\input{confidentiality-review}

\input{integrity-review}

\input{authentication-review}

\input{schema-randomization-review}


%% file: confidentiality-review.tex
\subsection{Confidentiality Aspect}

$Revision: 1.1.2.12 $

This section concerns with the confidentiality aspects
of data, i.e. data privacy and related issues of
searching, querying for statistical data, etc.

\subsubsection{Practical Encrypted Search}

If a database data resides on an untrusted server and, is, therefore, encrypted
for privacy considerations so the remote system administrator cannot snoop on
the data that is stored on that untrusted server. However, encrypting the database
complicates the search queries issued against that server resulting in a question
of how do we search inside the data that is encrypted, and yet, still maintaining
privacy of the data. This problem is first addressed in \cite{practical-encrypted-search},
from which we will re-cite the final solution scheme, Scheme IV, as in \xf{fig:encrypted-search-scheme}.

\begin{figure}
	\centering
	\includegraphics[width=.7\textwidth]{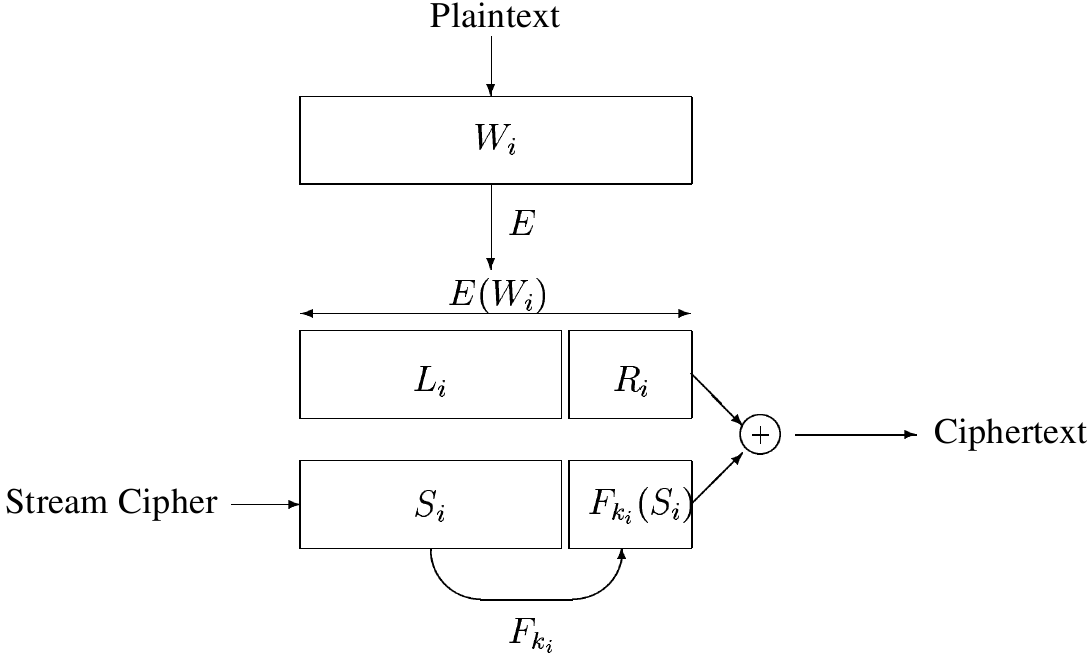}
	\caption{Scheme for Encrypted Search \cite{practical-encrypted-search}.}
	\label{fig:encrypted-search-scheme}
\end{figure}

This algorithm is provably secure and
cryptographically-efficient (no public key operation and
small message expansion). However,
there are open problems, such as searching with
ranged queries (e.g. ``$value > 13$'') or
regular expressions (``a[a-z]b'') as indexing and expansion
in the encrypted world is an issue. Such ranges and
expressions can sometimes be solved by ``flattening''
the expressions into $N$ equality queries, but this
may be a LOT of queries (e.g. in the latter case it
would be 26), which may unnecessarily provide additional statistics
for analysis to a potential intruder.

\subsubsection{$k$-Anonymity}

This brief summary is based on the paper and the course lecture notes \cite{kanonymity2002,wang-inse691a-2007}.
The basic idea of $k$-anonymity is to prevent linking
a record to an individual (in case of personal records of people),
e.g. ``that voice sample A belongs to Serguei Mokhov''. If a
record can be linked to $k$ or more individuals, that's an
acceptable risk in case we know a possible attacker can attack
by linking sanitized data from different sources, so we make
our $k \geq 2$. The problem with this approach is one cannot
possibly know all possible ways of linking attacks for it's
not feasible to account for or have a control over all possible
external data sources. One approach is to further modify the
data being released. The typical solution here includes
identifying so-called {\em quasi-identifiers} (attributes that
may be used to link), alter the table such that each combination
of the quasi-identifier values corresponds to at least $k$ records.
There are two main ways of altering the table: generalization
(releasing less accurate data, which is different from perturbation, e.g.
stripping off the day of a month from birth date), and suppression (when
some data is not released at all).

In generalization we distinguish {\em domain generalization} and
{\em value generalization}. A cross product of the two produces
a lattice structure corresponding to the different instances of
the private table and its generalized to a various degree instances
on one or more attributes.

$k$-minimal generalization indicates that a generalized table GT
satisfies $k$-anonymity and is minimal (i.e. no other generalized
table exists that can satisfy $k$-anonymity and be generalized
by GT at the same time).

Suppressing a record reduces the level of generalization, and,
therefore, provides better accuracy of released data. Given
a generalization level, {\em minimal required suppression} removes all
and only the records that fail $k$-anonymity requirements giving
the generalization higher priority.

The minimal $k$-generalization can be computed in a few ways.
In the naive approach we search for each {\em locally} minimal
generalization along each path in the table generalization hierarchy
bottom up. Then, we compare those locally minimal generalizations
and choose the globally preferred result. In the binary search
approach, if a generalization fails the $k$-anonymity criteria, then
those lower than it in the hierarchy will fail too. We do the binary
search with respect to the height, that is the distance from the bottom.

We can set up preference policies, such that
if the first hierarchy has e.g. 100 elements
while another one has two, then [1,0] may be much
better than [0,1].  As another example, the result requiring least
suppression may be desired over those with less
generalization.

\paragraph*{Cost of $k$-Anonymity}

The usability of a relation is reduced by generalization and
suppression.
Either way, some information is removed.
$k$-anonymity is obtained at a certain price, and
we'll use the discernibility metric (DM) to quantify
it:

\begin{enumerate}
\item
Each generalized record is assigned
a cost: how many records there are identical to it.
\item
Each suppressed tuple is assigned
a cost equal to the relation's original size before suppression.

\item
Thus, the discernibility metric (DM) is expressed as follows.
The relation $D$ is partitioned into $E_1$, $E_2$, ... $E_n$:
$DM = \sum_{|E| \geq k}|E|^2 + \sum_{|E| < k}|D||E|$

\item
NOTE: to satisfy $k$-anonymity with least cost is computationally infeasible.
\end{enumerate}

\subsubsection{$l$-Diversity}

$k$-anonymity does not guarantee privacy in itself \cite{wang-inse691a-2007}.
For example, a relation can be 2-anonymous
but the two individuals will be concerned if their
records are revealed if they have the same problem/disease.
Thus, we introduce $l$-diversity: each anonymized group
must have at least $l$ well-represented groups.
Simplest form: at least $l$ different values (e.g., $l=2$),
and if some of the records do no satisfy $l$-diversity, then
they may need to be suppressed when the data is released.

\subsubsection{$k$-Uncertainty}

$k$-anonymity makes a big assumption:
only a single relation is released \cite{wang-inse691a-2007}.
One has all the time to modify the table before
releasing it. A different problem arises:
how to achieve privacy guarantee if users can ask
database queries. This means, more than one relation may be released.
These relation are different views of the same original private relation,and
therefore, they are related.
Thus, there is a more complicated problem related to inference
of the private data.

The association between name and problem is no longer
as simple as ``George has either Cold or Obesity''
It could happen to be: ``George must have Cold, but may
or may not have Obesity''.

\paragraph*{Assumptions}

We have a single private table $Tbl$.
What is released:
view set $v$: a set of materialized views and
view definitions (i.e., the queries used to define the views).
In addition, the ``public'' knows the table schema.
(Notice: the problem will be very different, if the view
definitions and schema are not known.
In anther words, these definitions provide adversaries a lot more
information to make educated guesses).

%
%
%
%
%

Let $I^v$ be the collection of all possible database
instances that satisfy all the released views.
Define an association cover as:
a set of pairs $\{(ID,P)\}$, such that for any possible
database instance $I$ in $I^v$: $\Pi_{ID,P}(I) \bigcap C \neq \oslash$
Meaning, whatever the database is, it must contain
one of the pairs inside the association cover.
In other words, one of the pairs must be true.


In $k$-uncertainty
there are usually more than one association cover.
If the released views satisfy say 2-uncertainty,
an adversary must exclude one of the
possibilities before they can know anything for
sure.

\subsubsection{Indistinguishability}

Uncertainty is not enough for privacy \cite{wang-inse691a-2007}.
%
%
%
%
The possible privacy breach may result due to the difference
between individuals induced by the released views.
Violation a privacy requirement, the protection from being
brought to the attention of others.
We need to consider the other aspect, {\em indistinguishability}
where an individual is hidden in a crowd who have similar/same
possible private values.
In the notion of {\em symmetrical indistinguishability}
we have a metric $k$-SIND: for every possible instance, the
definition requires that two SIND B tuples can swap
their possible private values and still yield the view $v$.

\paragraph*{Complications}

Uncertainty by itself does not guarantee indistinguishability,
nor indistinguishability does not guarantee uncertainty.
The indistinguishability definition must hold for every possible
instance, not just for the actual instance, which has efficiency
implications when releasing data.

\subsubsection{Parameters Summary}

\begin{enumerate}
\item Data object to anonymize (encrypt).
\item Encryption key(s) and their size.
\item Encryption algorithm type (CBC-DES, RSA, DSA).
\item Hashing algorithms for HMAC (SHA1, MD5) for encrypted search.
\item $n$ is the size of the search word, and $m$ is the size of the right
portion in bits that corresponds to the encrypted portion of $W_i$ on the right $R_i$
and the same size as $F_{k_i}(S_i)$.
\item $k$ -- an integer, how many records should appear similar at a minimum, $\geq 2$.
or how many association there may be for the $k$-uncertainty, or the parameter for
undistinguishability $k$-SIND.
\item Confidentiality algorithm type ($k$-anonymity and/or $l$-diversity and/or $k$-uncertainty) with ability to chain
the algorithm or set any combination of them depending on the desired policy.
\item $l$ -- an integer for $l$-diversity.
\end{enumerate}


%% file: integrity-review.tex
\subsection{Integrity Aspect}
\label{sect:integrity}

$Revision: 1.1.2.22 $

This section covers the database integrity aspects, primarily
borrowed from these works:
\cite{nasiro02-digital,leandro02-audio,hsien05-image,rakesh02-database,rabu04-database}.

\subsubsection{Integrity Lock Architecture}

The integrity lock architecture is an architecture which enforces integrity check by insertion of
timestamps in (e.g. watermarking) or append stamps on (e.g. data authentication) in the data.
A trusted front-end must enforce MAC function by filtering out disallowed data for select queries based on stamps.
It also controls the queries by updating and adding stamps in or append to the data.
The checksums (or watermark) help to verify their integrity. Since in our
project we are not dealing with users or multiple levels of access control,
we simplify the integrity lock architecture as shown in \xf{fig:marf-security-integrity-integritylock}.

\begin{figure}
	\centering
	\includegraphics[width=.5\textwidth]{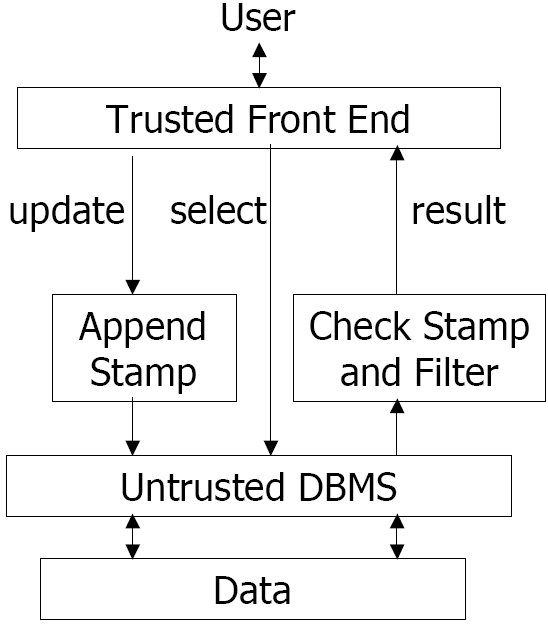}
	\caption{Integrity Lock Architecture}
	\label{fig:marf-security-integrity-integritylock}
\end{figure}

\clearpage

\subsubsection{Introduction to Watermarking for Integrity Checks}

For the assurance of database integrity, we can use authentication mechanism (see \xs{sect:authentication}) with
cryptographic methods to generate an authentication signature with a key and prove the data
origin by verifying with the same key.
However, the authentication mechanism like signature to secure integrity has some drawbacks \cite{wang-inse691a-2007}:

\begin{itemize}
\item
Large overhead, can't be used on a per item basis
\item
If used on relations, can't localize the modification.
Once the signatures mismatch, the entire relation is useless.
\item
The signatures must be stored elsewhere.
\end{itemize}
However, the watermarking technologies come in handy to try to address these problems.
The watermarking technique has been implemented for multimedia for decades.
It is used in multimedia products to prove the ownership of copyright or prevent piracy.
Some research has been done for watermarking of databases.
Moreover, more and more databases store not only primitive data types, but also multimedia data, like audio or video data.

Now, the researchers in the digital watermarking field have also been thinking to use
it for data stream or even database integrity checking.
Therefore, we would like to focus on digital watermark technologies for multimedia with
database integrity and their implementation.

Multimedia copyright enforcement is not a new issue. The recording industry has been fighting piracy since its very early times.
However, the digital revolution in audio, image and video has brought this fight to a new level, as all the multimedia in digital format can be copied and distributed easily and with no degradation.
Electronic distribution, particularly the Internet, associated with efficient compression algorithms (such as AVI, MPEG, WAV and MP3) and peer-to-peer file-sharing systems (such as Emule, LameWire, Napster and Gnutella) create an environment that is propitious to multimedia piracy.
Watermarking has been proposed as a potential solution to this problem.
It consists in embedding a mark into the original media signal, which is different from signature, and attaches to the original data or is stored separately.
This mark should not degrade media quality.
In order to prove the origin of content, the mark should be detectable and retrievable.
In our report, we would like to introduce the three types of media which are audio, image, and video.

\subsubsection*{Digital Watermarking Requirements}

Digital watermarking is a kind of process to inject a watermark into media object without reducing the quality of it self.
Therefore, the watermark can be extracted and used to be a proof the origin of source media.
It also can be evidence while being used in law suit.
The seller of the digital data knows one of his videos has been edited and published without payment of royalties.
Therefore, the detection of the seller's watermark in the digital data can be served as evidence that the published video is property of that seller.
The watermark used for media should achieve four components of requirements which are invisibility, robustness, security, and capacity.
Without any one of these requirements, the watermark will be useless to protect integrity of original data.

\begin{itemize}
\item
Invisibility: for any multimedia, the watermark must be invisible; otherwise, it will reduce the value of the multimedia object.
If music comes with a watermark noise, then it will affect the quality of the music.
\item
Robustness: for any multimedia, it is so easy to find software to rip from original media, for example, a DVD movie.
The quality of multimedia can be modified to fit the users need. Therefore, in order to prove multimedia objects, the watermark is supposed to keep some kind of readable quality even if the user modifies the content of a multimedia object.
Then, the watermark can be an evidence in the court to prove the object belongs to the original owner.
\item
Security: the watermark shouldn't be removable. The watermarking technique should use a cryptographic algorithm to inject the watermark inside of a multimedia object such that there is no way to remove or modify it.
In other words, it should be impossible to reverse the injection procedure with the purpose of removal of the watermark without knowing the secret information (like secret keys).
\item
Capacity: the watermark should be easy and fast to embed in the multimedia object. It shouldn't take very complicate or long process to do it.
\end{itemize}

\subsubsection*{Functionality of Watermarking}

Watermarking can be used in many kind of purposes.
It also offers tremendous functionality for application.
The functionality of watermarking technique can be summarized below:

\begin{enumerate}
\item
Ownership assertion. Watermarks can be used for ownership assertion.
To assert ownership of an image, owner can generate a watermarking by a secret key and then embed it in original media or data by watermarking algorithm.
He can then make the watermarked image available in public.
Later, if a forger has the ownership of an image which is from this public image, the original owner can present the original object and retrieve the watermark from the forger's.
For such scheme to work, the watermark should be robust enough to prevent malicious removal.
In addition, the watermark should be impossible to forge.
\item
Fingerprinting. If the multimedia content is electronically distributed over the Internet, the owner may like to prevent unauthorized duplication by embedding a distinct watermark (or a fingerprint) in each copy of the data.
If, unauthorized copies of the data are found, then the origin of the copy can be determined by retrieving the watermarking (or fingerprint). In this case, the watermark needs to be invisible and must also be invulnerable to be forged, removed or invalidated.
Furthermore, the watermark should be resist to collusion. That is, a group of $k$ users with the same image but containing different fingerprints should not be able to collude and invalidate any fingerprint or create a copy without any fingerprint.
\item
Copy prevention or control. Watermarks can also be used for copy prevention and control. A digital watermark can be inserted and indicate the number of copies that are permitted.
Every time a copy is made the watermark can be modified by the hardware.
An example of such a system is the Digital Versatile Disc (DVD).
In fact, a copy protection mechanism that includes digital watermarking may include the ability to read watermarks and react by their presence or absence. Another example is in video frame.
The information can be embedded as a watermark in every frame or a sequence of frames to help investigators locate the piracy more quickly.
\item
Fraud and tamper detection. For multimedia content is used for legal purposes, medical applications, news reporting, and commercial transactions, it is important to ensure that the content was originated from its source and that it had not been changed or falsified.
This can be achieved by embedding a watermark in the data. The watermark can also include information from the original image that can help to recover any modification.
On the other hand, for watermark been used for authentication purposes, it should not affect the quality of an image and should be resistant to modification.
\item
ID card security. Information in a passport or ID can also be included in the owner's photo that appears on the ID. The ID card can be verified by the embedded information and written text on the ID.
However, the watermark can provide an additional level of security in this application. For example, if the ID card is forged or stolen, then the failure in extracting the watermark will refuse the ID card.
\end{enumerate}

The above represent a few example applications where digital watermarks could potentially be of use.
In addition, there are many other applications in copyrights management and protection
like tracking use of content, monitoring broadcasting, binding content to specific players.
Therefore, a digital watermarking technique needs to satisfy a number of requirements.
Since the specific requirements vary with the application, watermarking techniques need to be designed within the context of the entire system in which they are to be injected.
Each application implies different requirements and would require different types of watermarking schemes or a combination thereof.
For the rest of report, we would discuss different watermarking principles and techniques for different type of data.
Our goal is to present a better understanding of the basic principles of digital watermarking.
We will focus on multimedia and relational database watermarking in our discussions and examples.

\subsubsection{Audio Watermarking}

As we know, audio signal is kind of wave signal. Therefore, the basic idea consists in how to add an audio signal, the watermark, to the original audio signal.
The watermarked signal must be only minor distorture and perceived by the listener as identical to the original one.
The watermark carries data that can be retrieved by a detector and can be used for a multitude of purposes.

\paragraph*{Audio Watermarking Requirements}

The requirements that an audio watermarking system must satisfy
are application-dependent and we can mention as general requirement:

\begin{enumerate}
\item
Inaudibility: the audio watermarking shouldn't be perceived by the listener and should not degrade sound quality.
\item
Robustness: audio watermark should resist any transformations applied to the audio signal, and sound quality is not unacceptably degraded by modification.
\item
Capacity: audio watermark bit rate should be high enough for the application to capture, which must be balanced with inaudibility and robustness; a trade-off must be defined.
\item
Low complexity: for real-time applications, watermarking algorithms should be acceptable time-wise.
Some applications (such as low bit-rate audio over the Internet) might admit the watermark to introduce a small level of sound quality degradation, while others (such as high bit-rate audio) would be extremely rigorous.
\item
Reliability: the watermark must allow some portion of error bits. Data contained in the watermark should be extracted with acceptable error rates.
\item
Resistance: to signal-processing operations such as resampling or filtering is usually necessary for resistance.
For copyright protection, resistance to malicious attacks target on preventing watermark detection is also required;
for example, if a piece of the signal is deleted, the watermark should still be detectable.
On the contrary, for integrity-verification applications (such as tape of testimonies presented in the court),
the watermark must be weak, fragile, and no longer be recognized once the audio is modified by unauthorized people.
\end{enumerate}

The audio watermarking can be treated as a communication system.
The audio signal carrying useful information and channel noise.
In traditional communication systems, the useful signal is usually much stronger than the noise, and the noise is often assumed to be Gaussian and white.
To avoid audible distortion, the watermark signal must be much weaker than the audio signal. Furthermore, the audio signal is generally non-stationary.
Some basic approaches for audio watermarking have been proposed in the research. For example, we can mention:

\begin{itemize}
\item
Spread-spectrum watermarking: As in spread-spectrum communication systems (Dixon, 1976) (Haykin, 1988)
the idea consists in spreading the watermark in frequency to maximize its power while keeping it inaudible and increasing its resistance to attacks (Boney et al.,1996) (Garcia, 1999).
\item
Echo-hiding watermarking: Temporal masking properties are exploited in order to render the watermark inaudible.
The watermark is an echo of the original signal (Bender et al., 1996) (Neubauer, 2000).

\item
Bit stream watermarking: The watermark is inserted directly in the bit stream generated by an audio coder. For example, in (Lacy et al., 1998), the watermark consists in the modification of scale factors in the MPEG AAC bit stream.
\end{itemize}

\paragraph*{Concept of Masked Watermark}

Psychoacoustics is the study of the perception of sound.
The study is, when two tones is close to each other in frequency and they are played simultaneously,
then ``frequency masking'' happen: if one of the tones is sufficiently loud, it masks the other one (Zwicker \& Fastl, 1990).
Psychoacoustic models generalize the frequency-masking effect to non-tonal signals.
In audio watermarking, psychoacoustic models are often used to ensure inaudibility of the watermark.
The watermark is constructed by inserting in frequency a nearly-white signal according to the masking threshold.
After this operation, the watermark is always below the masking threshold and the watermark shouldn't heard in the original sound signal. In \xf{fig:marf-security-integrity-audiofrequencymasking} is the example of such an encoding.

\begin{figure}
	\centering
	\includegraphics[width=.6\textwidth]{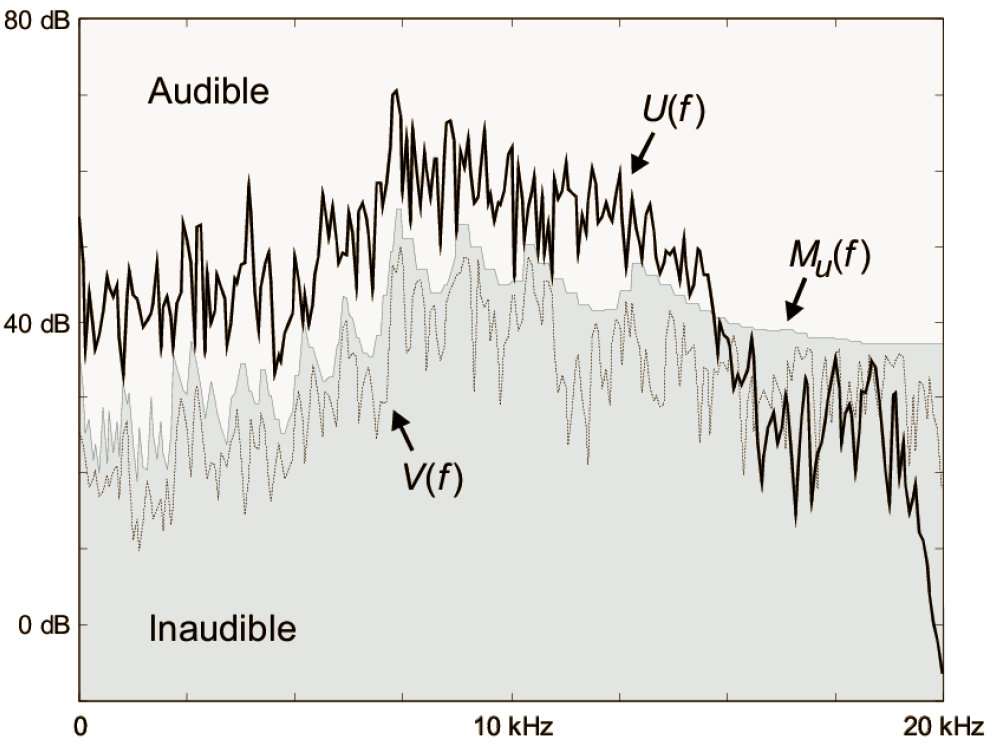}
	\caption{Audio Frequency Masking Algorithm}
	\label{fig:marf-security-integrity-audiofrequencymasking}
\end{figure}

\subsubsection{Image Watermarking}

The digital image, has same problem as digital audio, can be manipulated using variety of sophisticated image processing tools easily.
Such manipulations are so easy and emphasize the need for image authentication/verification techniques in applications.
Basically, image watermarking has same requirement as audio watermarking, which are invisibility, robustness, security and capacity, and same functionalities. Therefore, in this part, we would like to introduce a basic algorithm for image watermarking.
The image watermarking scheme is the same as audio watermarking scheme.
However, we know that audio signal is different from image signal by their domain, same as watermarks for each.
Audio watermarking focus on audio wave frequency domain, on the other hand, image watermarking focus on spatial or color frequency domain. Therefore, for the image watermark injection must have different approaches to work on.

\paragraph*{Background}

The image watermarking schemes are classified into the methods of the spatial domain and the frequency domain. To hide a watermark in the frequency domain, an image has to be transformed from a spatial domain into its frequency domain.
This scheme requires many computations and time to embed or retrieve the watermarks. Meanwhile in the spatial domain, the watermark can be directly embedded into the pixels values. The algorithms for embedding and recovering are simple.
Traditionally, the scheme hides the watermark bits in the least significant bits (LSB), which should make imperception for color manipulation for viewer.
Several techniques can be used for a image transform (e.g. discrete Fourier, discrete cosine, Mellin-Fourier, Wavelet). Then, insert the watermark in the transformed space. last, invert the transform back to get original marked image.
The noise caused by the watermarking signal is thus spread over the whole image without being visible.

\paragraph*{Wong's Image Authentication Watermark}

We demonstrate the Wong's scheme is a block-based watermarking technique.
In his scheme, he gives an $M x N$ image $X$, a binary watermark image $W$ of the same size as the original.
In practice, this step is usually achieved by tiling the original image with a smaller logo image.
The original image $X$ is partitioned into $O$ x $P$ pixel blocks, \{ $X1, X2$, ...\}; where $X_{r}$ denotes such blocks.
The watermark image is partitioned into blocks $W_{r}$.
For each block $X_{r}$, a corresponding block $\tilde{X_{r}}$ is formed by setting the least significant bit of each pixel to zero.
A cryptographic hash, e.g., MD5 or SHA1, can be used to transform block $\tilde{X_{r}}$ and image dimensions is computed:

$H_{r}=\mathit{H}(M,N,\tilde{X_{r}})$

The hash value $H_{r}$ can be treated as a random number.
To get the signature, we compute as below:

$S_{r}=Encrypt(H_{r} \oplus W_{r}, Key_{private})$

In the last step, we insert the signature $S_{r}$ into $X_{r}$ block as least significant bits of the block.
The key point for this algorithm is the watermarking insertion is independent on each block.
The watermar verification is similar to watermarking authentication.
First, we partition image $\hat{X}$ into blocks $\hat{X_{r}}$.
From LSB of the $\hat{X_{r}}$ block, we get a signature $\hat{S_{r}}$.
Lastly, we get $H_{r}$ by setting LSB to zero and calculate:

$\hat{W_{r}}=Decrypt(\hat{S_{r}}, Key_{private}) \oplus \hat{H_{r}}$

Any change in the pixel values in each block modify decrypted signature or hash value.
Therefore, any modification can be detected and located in the corresponding image block.

\subsubsection{Video Watermarking}

\paragraph*{Video Watermarking schemes}

Video constructs by frames (images).
Therefore, it has same structure of image, but images. However, the video watermarking have much more problems to be concerned,
like frame shift, frame dropping, cropping, scaling, rotation, and change of aspect ratio, especially when some of these are combined together.

The requirement of video watermarking is the same as image watermarking. However, the video watermarking techniques have much more sophisticated schemes than audio and image watermarking schemes do.
There are so many schemes proposed in public, like KLT (Karhunen Loeve transform), DCT (discrete cosine transform), DFT (discrete Fourier transform), DWT (discrete Wavelet transform), WPT (wavelet packet transform).
In our paper, we would like to shortly introduce three video watermarking schemes, which are DWT-based Blind Video Watermarking Scheme, Multiresolution Scene-based video watermarking and Robust MPEG Video Watermarking Using Tensor:

\begin{description}
\item [DWT-based Blind Video Watermarking Scheme]:
The Discrete Wavelet Transform watermarking scheme is used to transform video frames to wavelet domain, then inject watermark (it may be a logo or trademark) in to frames of each scene. After that, the application can perform scene change detection.
Since each scene is embedded with a same watermark, so it can prevent attackers from removing the watermark by frame dropping. Independent watermark used for successive different scene can prevent attackers from using the frames completely different scenes.

\item
[Multiresolution Scene-based video watermarking]:
This algorithm still uses the Discrete Wavelet Transform method, but it filter low-pass frames (static, non-moving component) and high-pass frames (dynamic, moving component) into different blocks.
However, the problem of this scheme is that the frequency masking is tuned to human visual perception.

\item
[Robust MPEG Video Watermarking Using Tensor SVD]:
Unlike previous methods where each video frame was marked separately, this method uses high-order tensor decomposition of three dimensional video scenes.
The Key idea behind this methodology is to utilize a fixed number of the I-frames as a multidimensional tensor with two dimensions in space and one in time.
Next, transform a tensor into a matrix. This process called ``matricizing''.
The advantage of this scheme is the extracted watermark from the modified video is easy to verify its watermark.
This method is proved robustness for video cropping, dropping or even altering attack.
\end{description}

\subsubsection{Database Watermarking}

The database watermarking is a new technique where one can inject watermark into a database, then to authenticate and verify by retrieving the watermark later.
The watermarking software represents some small error bits into the database object being watermarked.
These intentional errors are called marks and all the marks together construct the watermark.
The marks must not have a significant impact on the usefulness of the data and they should be placed in such a way that a malicious user cannot destroy them without making the data less useful. Thus, watermarking does not prevent copying,
but it thwarts illegal copying by providing a means for establishing the original ownership of a copied data.

\subsubsection*{Fragile Watermarking}

Generally, the digital watermarking for integrity verification is called fragile watermarking as compared to robust watermarking for copyright protection.
In a robust watermarking scheme, the embedded watermark should be robust against attacks from removing or invalidating the watermark.
However, the fragile watermarking scheme imply that the watermark should be fragile to modifications.
Once been modified, the application can detect and localize the modifications.
For example, integrity of an database record can be controlled by means of a fragile watermark. If the watermarked
record is edited, the watermark can not be verified, but the application can localize the modification. On the other hand, in relational database, any un-authorized
modification can corrupt the trust of a specific range of tuples which hides a watermark inside.

\subsubsection*{Why Database Watermarking is Standing Out}

The multimedia watermarking technique discussed earlier always have same domains to focus on.
For example, audio frequency domain, pixel frequency domain, spatial/temporal domain.
However, databases don't have such patterns to follow.
Therefore, we would discuss the difference between multimedia and database watermarking techniques:

\begin{itemize}
\item
As we know the multimedia object consists of a large number of bits with considerable redundancy.
Therefore, the watermarking can have a large of room to hide. However, a database consists of attributes, tuples, and relations.
Each of them are objects and most or all bits of objects are likely meaningful.
The watermarking method tries to insert watermark in these separate objects.

\item
For media files, they all have frequency or spatial/temporal domain and these attributes don't change in different media files. However, attributes consists different types of primitive objects and tuples in a relation constitute a set of attributes and there are no specific rules to follow.

\item
Portions of a multimedia object cannot be dropped or replaced without causing perceptual changes in the object. However, the modification of a relation can simply drop some tuples or substitute them with tuples from other relations without notice.

\item
For insertion of watermark in image or video, we use the techniques to transform image by discrete Fourier, discrete cosine, Mellin-Fourier, and wavelet methods. However, if we apply these techniques to a database will produce errors in all of the attribute values, which might not be acceptable.
Furthermore, such a watermark might not survive even minor updates to the relation.
\end{itemize}

\subsubsection*{Database Watermarking Method}

As mentioned before, the sensitive data may be changed and numeric may transformed. Therefore, it is important to decide the watermarking method for different database scheme. Here, we would like to discuss watermarking technique of relational database.
The relational database watermarking is a resilient watermarking method for relational data.
It is a technique for enabling user-level runtime control over properties that are to be preserved as well as the degree of change introduced,a complete, user-friendly implementation for numeric relational data.

\paragraph*{Relational Database Watermarking}

We would like to discuss the relational database watermarking method which is injected watermark in tuples. The technique marks only numeric attributes. One of most important assumption is the marked attributes are made such a small change. This change can be ignored by data owners.

\paragraph*{Algorithm of Relational Database Watermarking}

In order to describe the algorithm, we define all the notations in the beginning:

\begin{itemize}
\item
$\eta$	--	Number of records in the relation
\item
$\nu$	--		Number of attributes in the relation available for marking
\item
$\xi$	--		Number of least significant bits available for marking in an attribute
\item
$1/\gamma$	-- Fraction of records marked
\item
$\omega$	--	Number of records marked
\item
$\alpha$	--	Significance level of the test for detecting a watermark
\item
$\tau$	--	Minimum number of correctly marked tuples needed for detection
\end{itemize}

For watermarking of database relation $R$ whose scheme is
$R(P,A_{0},...,A_{\nu-1})$, where $P$ is the primary key attribute.
For simplicity, assume that all $\nu$ attributes $A_{0},...,A_{\nu-1}$ are candidates for marking.
They are all numeric attributes and their values are such that changes in $\xi$ least significant bits for all of them are imperceptible.
Now, we present the watermark insertion algorithm in \xf{fig:marf-security-integrity-watermarkinsertion}.

\begin{figure}
	\centering
	\includegraphics[width=.6\textwidth]{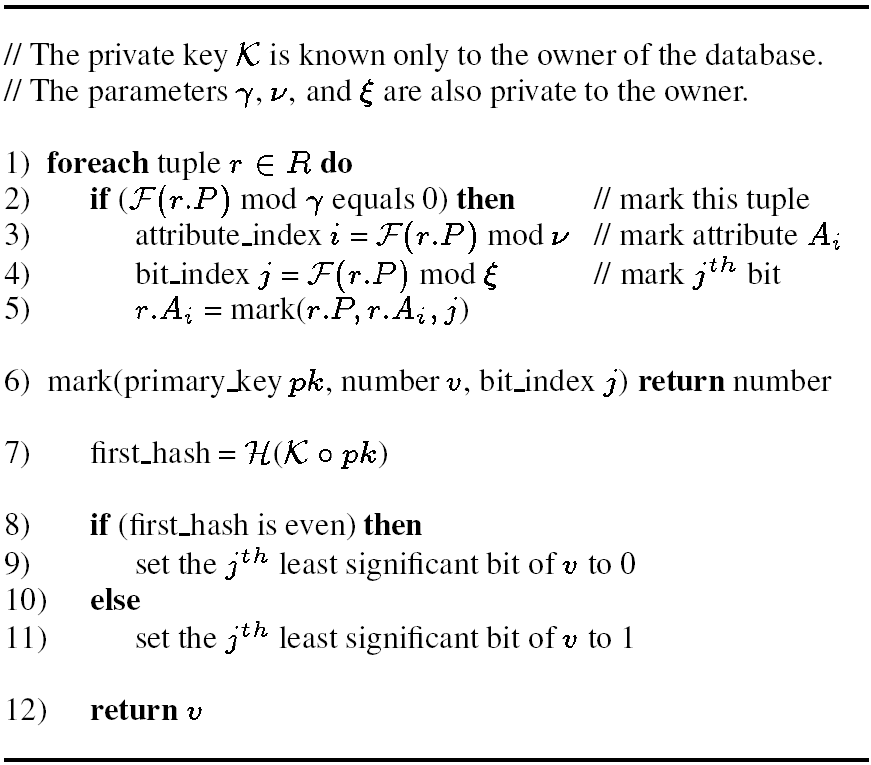}
	\caption{Watermark Insertion Algorithm}
	\label{fig:marf-security-integrity-watermarkinsertion}
\end{figure}

In this algorithm, there is a function $F(r.P)$ is calculated by below formula.

$F(r.P) = \mathcal{H}(\mathit{K} \circ \mathcal{H}(\mathit{K} \circ r.P))$

The $\mathcal{H}(M)$ represents hash of the message $M$.
The $\circ$ represents concatenation.
Next, we present the watermark detection algorithm to retrieve watermark from rational database
in \xf{fig:marf-security-integrity-watermarkdetection}.

\begin{figure}
	\centering
	\includegraphics[width=.6\textwidth]{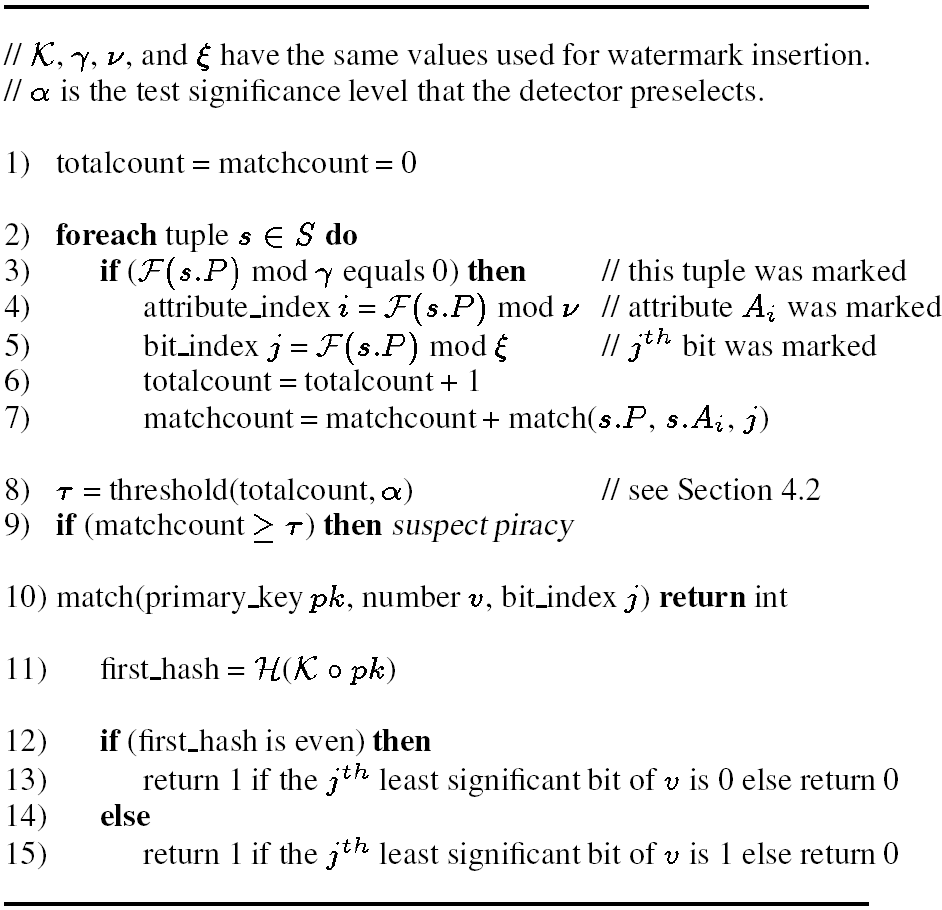}
	\caption{Watermark Detection Algorithm}
	\label{fig:marf-security-integrity-watermarkdetection}
\end{figure}

\subsubsection{Parameters Summary}

For the parameters of implementation, we present the parameters of relational database watermarking.

\begin{enumerate}
\item
	Number of records in the relation
\item
	Number of attributes in the relation available for marking
\item
	Number of least significant bits available for marking in an attribute
\item
	Fraction of tuples marked
\item
	Number of tuples marked
\item
	Significance level of the test for detecting a watermark
\item
	Minimum number of correctly marked tuples needed for detection
\item
	A private key known only by the owner
\item
	The type of MAC hash/timestamp function for integrity lock and its key if is a keyed function

\end{enumerate}

\subsubsection{Discussion}

The watermarking is a mechanism which can provide strong data integrity assurance for multimedia data or even a relational database. With embedded watermark, the data or relational database can assure data origin and prevent from malicious modification.

%% file: authentication-review.tex
\subsection{Authentication Aspect}
\label{sect:authentication}

$Revision: 1.1.2.10 $


This section presents the shortened summary of the research done on the authentication aspects of various types of data, such
that the framework being developed pragmatically covers most aspects and parameters to be flexible and uniform.
%
%
The data authentication aspect in {\jdsf} is mostly about data-origin's authentication
(e.g. in {\dmarf}~\cite{dmarf06} or the General Intensional Programming System ({\gipsy})~\cite{gipsy} the
data can easily come from another host during the distributed computation that may have been spoofed
and is intentionally producing incorrect results passing them off as integrity-correct).
This review is primarily based on the cited
works~\cite{mnt_ndss_2004,dasfaa_2006,BUCS-TR-2006-011,sigmod06-btree}
as well as the lecture notes~\cite{wang-inse691a-2007}
and some related techniques discussed in our integrity
work~\cite{jdsf-integrity-cisse08}.

\subsubsection{Authentication Introduction}

Any type of storage management system, such as DBMS, etc. is important in the vast number of applications.
It is a problem, when it involves the data owners
delegating their data management needs to
an external service provider (e.g. in our simple example for {\marf} to {\hsqldb}).
Since a service provider most of the time is not
fully trusted from a variety of security aspects point of view, 
there is one of the several core security requirements we study
is the 
authenticity of the outsourced data and databases.
Outsourced databases (ODB) is a relatively recent paradigm that has been proposed and received considerable attention.

There is still a lot research to do to develop the ODBs to be fully trusted.
The basic idea is that data owners delegate their database needs and functionalities to a
third-party storage provider, which offers services to the users of the database.
Since the third party can be untrusted or can be compromised, security concerns must
be addressed before this delegation takes place.

The database outsourcing paradigm poses numerous research and development challenges,
which do not affect the overall performance, usability, and scalability,
but impact one of the foremost challenges that is the security of stored or transmitted data.
For example, a user stores their data
(which are usually a critical asset or confidential matters)
at an external, and potentially untrusted,
data storage service provider.
It is thus important to secure the outsourced data from potential attacks
not only by malicious outsiders but also from the service provider itself.
Consequently, whenever the users try to query from a hosted database,
the results must be demonstrably authentic (with respect to the actual data owner)
to make sure that the data came from a legitimate source 
(and also have not been tampered with, which the integrity aspect assures).

Thus, 
we focus on researching to provide secure and effective means of ensuring data authentication,
while incurring minimal computational and bandwidth overhead.
In particular, we investigate techniques to help the ODB clients to authenticate
the origin of data coming from the service provider the data owner as a query.
At the end of this section we summarize a few solutions, which have been researched
and published on how to authenticate data.
The goal is to design and implement these methods on top the existing
platforms of {\marf} and {\hsqldb} and beyond in the uniform manner,
which at this point do not have data authentication system built-in.

\subsubsection{Authentication Scope}

In the non-relational world (Java object serialization, XML, CSV, etc.)
and equivalent read/write queries have to be authenticated,
to make sure the underlying store was not swapped underneath
a running application (while its integrity may still be correct, but the
data may no longer be authentic), but that comes from an unauthorized provider
(techniques similar to those that can be borrowed from the
DNSsec~\cite{dnssec,rfc3833,rfc4034,rfc3225}
for host authentication).
In relational databases, we are to consider only the equality
queries and also logical comparison predicate clauses.
In other words, one considers the standard SQL queries involving
\api{SELECT}-type of clauses, which typically result in selection
of a set of records (or attributes) matching a given predicate or a set thereof.
In other hand, we do not consider queries that involve any kind of data aggregation for example \api{SUM} or \api{AVERAGE}.
We focus on the mechanisms for origin authenticity of query replies
returned by the storage service provider in the ODB model.
Another issue, which is equally important, is the completeness
of query replies that we consider in our integrity work~\cite{jdsf-integrity-cisse08}.

One of the existing solutions is the owner creates a specialized data structure over the original
database that is stored at the servers together with the database.
The structure is used by a server to provide a verification object $VO$ along with the answers,
which the client can use for authenticating the results. In our framework
design the notion of $VO$ is realized in the \api{AuthenticatedObject} class shown in \xf{fig:marf-security-storage}.
Verification usually occurs by the means of using classical collision-resistant hash functions and
digital signature schemes.
Note that in any solution, some information that is authentic to the owner
must be made available to the client,
and from the client's point of view,
the owner cannot be differentiated from a (potentially malicious) server.
Examples of such information include the owner's public signature
verification key or a token that in some way authenticates the database.
Any successful scheme must make it computationally infeasible for a malicious
server to send incorrect query results and verification object that will be accepted by a
client who has the appropriate authentication information from the owner.

\subsubsection{Cryptography Essentials}

The classical digital authentication algorithms involve cryptographic
signatures and hashing functions as well as more advanced data structures.
Due to shortage of space and the abundance of the general knowledge of them, we mention
them only briefly in this work.

\paragraph*{Collision-resistant Hash Functions}

A hash function takes a variable-length input and produces a fixed-length output $y=\mathcal{H}(x)$.
This creates a possibility of collisions (two or more distinct documents might map to the same hash value).
Such functions are collision-resistant if it is difficult to find such useful from the attacker's
point of view documents to match the same hash value.
However, computing a collision resistance flaw is in general computationally infeasible.
In our work, we will be providing the components to allow
heuristic hash functions, which have the advantage of being very fast to evaluate,
as well as any other hash function implementations there may be, i.e. our authentication
framework does not discriminate between algorithms and allows researches to implement
anything they need for comparative studies or the actual application use.
As an example, a basic HMAC-based authentication (also good for the integrity checks~\cite{jdsf-integrity-cisse08})
is illustrated in \xf{fig:hash-function} and \xf{fig:hmac-auth-process}.
The hash functions at option are implemented using whatever algorithm implementation is available,
e.g. MD5, SHA1, and others.

\begin{figure}[htb!]
	\centering
	\includegraphics[width=.3\textwidth]{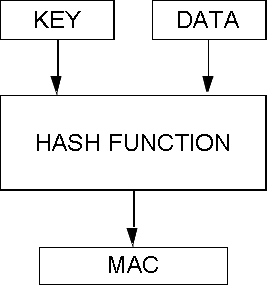}
	\caption{Basic HMAC-based Authentication \cite{assi-inse7120-2007}.}
	\label{fig:hash-function}
\end{figure}

\begin{figure}[htb!]
	\centering
	\includegraphics[width=.5\textwidth]{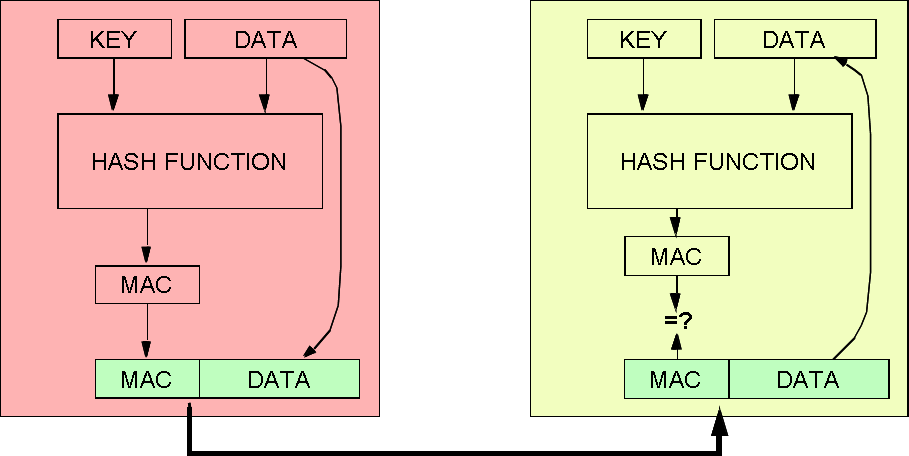}
	\caption{HMAC-based Authentication Process \cite{assi-inse7120-2007}.}
	\label{fig:hmac-auth-process}
\end{figure}

\paragraph*{Public-key Digital Signature Schemes}

A public-key digital signature scheme is a methodology that can be used
for authentication of both the integrity and ownership of a signed message.
In such a scheme, the signer generates a pair of keys -- a public, $k_{+}$,
and a private $k_{-}$, and the private key is used for data signing.
The classical digital signature algorithms include but not limited to
RSA, DSA, and ElGamel. For the large volumes of data, e.g. multimedia
data or large relational databases, such digital signature schemes alone are
computationally quite expensive, especially if applied per record.

\paragraph*{The Merkle Hash Tree}

The Merkle hash tree~\cite{merkle88,merkle90} is an improvement on solution for authenticating a set of data values.
It will solve the simplest form of the query authentication problem for point queries and datasets that can in main memory.
The Merkle hash tree is a binary tree, where each leaf contains the hash of a data value, and each internal node contains the hash of the concatenation of its two children.
The verification of data values is based on the fact that the hash value of the root of the tree is authentically published (authenticity can be established by a digital signature).
To prove the authenticity of any data value, all the prover has to do is to provide the verifier, in addition to the data value itself, with the values stored in the siblings of the path that leads from the root of the tree to that value.
The verifier, by iteratively computing all the appropriate hashes up the tree, at the end can simply check if the hash they have computed for the root matches the authentically published value.
The security of the Merkle hash tree is based on the
collision-resistance of the hash function used: it is computationally infeasible for a malicious prover to fake a data value, since this would require a hash collision somewhere in the tree (because the root remains the same and the leaf is different hence, there must be a collision somewhere in between).
Thus, the authenticity of any one of $n$ data values can be proven at the cost of providing and computing $\log_{2}(n)$ hash values, which is generally much cheaper than storing and verifying one digital signature per data value.
Furthermore, the relative position (leaf number) of any of the data values within the tree is authenticated along with the value itself.

\subsubsection{Solutions}

In this section we consider the solution to authentication problem in two case static case and dynamic case.

\paragraph*{Static Case}

In the static case, once the owner has initially created the database and published it to the servers there are no or very few updates in the system. In \xf{fig:marf-security-authentication-staticcase} is the sample high level code for this.

\begin{figure}
	\centering
	\includegraphics[width=.6\textwidth]{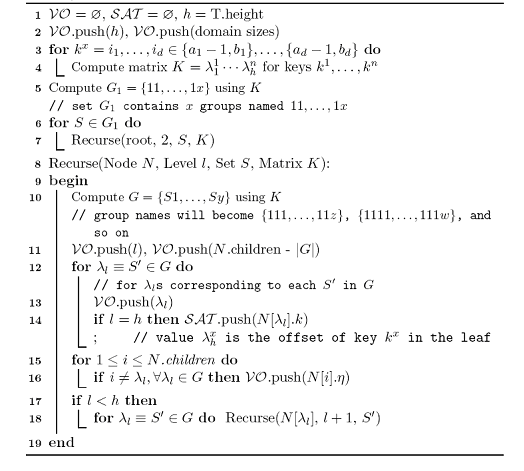}
	\caption{Authentication: Static Case}
	\label{fig:marf-security-authentication-staticcase}
\end{figure}

\paragraph*{Dynamic Case}

The APS-tree is a good solution for non-sparse, static environments because it has very small querying cost.
It will not work well though for dynamic settings.
In the worst case, updating a single tuple in the database might necessitate updating the whole tree.
This section creates advanced structures that overcome this limitation. The corresponding algorithms
are presented in \xf{fig:marf-security-authentication-dynamiccase}
and \xf{fig:marf-security-authentication-dynamiccase-verification}.

\begin{figure}
	\centering
	\includegraphics[width=.6\textwidth]{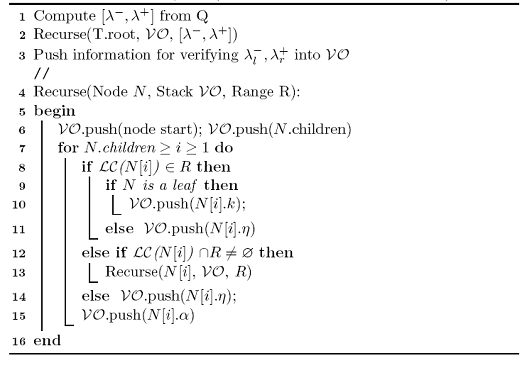}
	\caption{Authentication: Dynamic Case}
	\label{fig:marf-security-authentication-dynamiccase}
\end{figure}

\begin{figure}
	\centering
	\includegraphics[width=.6\textwidth]{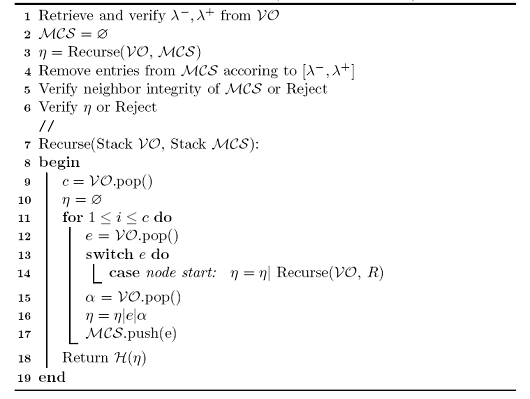}
	\caption{Authentication: Dynamic Case Verification}
	\label{fig:marf-security-authentication-dynamiccase-verification}
\end{figure}

\subsubsection{Aggregates}

In case of relational databases, \api{COUNT} is a special case of \api{SUM} and is thus handled similarly.
The combination of \api{SUM} and \api{COUNT} provides a solution for \api{AVG} as well.
AAB-tree and AAR-tree can be modified to support \api{MIN} and \api{MAX} queries,
simply by replacing the aggregate values stored in the index nodes of the trees,
with the \api{MIN}/\api{MAX} of their children. The APS-tree cannot handle \api{MIN}/\api{MAX} aggregates.
Authentication of holistic aggregates, like \api{MEDIAN}, is much harder and left as future work.

\subsubsection{Parameters Summary}

\begin{enumerate}
\item
	AAB-tree -- is an extended B+-tree structure
\item
	$VO$ -- Verification Object that contains the hashes stored
\item
	$SAT(Q)$ -- the set of records from $T$ that satisfy all query predicates
\item
 	$ANS(Q)$ -- the final answer to a query $Q$
\item
	$Q$ -- the aggregation query
\item
	$k$ -- Values that need to be authenticated.
\item
	$\mathcal{H}$ -- The type of MAC hash function
\end{enumerate}

%% file: schema-randomization-review.tex
\subsection{SQL/Schema Randomization Aspect}
\label{sect:sql-randomization}

$Revision: 1.1.2.12 $

The SQL/schema randomization aspects described in this section are
primarily derived from \cite{sqlrand-boyd-keromytis}.

\subsubsection{Introduction}

We present a practical protection mechanism against SQL injection attacks. Such attacks
target databases that are accessible through a web front-end, and take advantage of flaws
in the input validation logic of Web components such as CGI scripts. We apply the
concept of instruction-set randomization to SQL, creating instances of the language that
are unpredictable to the attacker. Queries injected by the attacker will be caught and
terminated by the database parser. We show how to use this technique with the an SQL-based
database using an intermediary proxy that translates the random SQL to its standard
language. Our mechanism imposes negligible performance overhead to query processing
and can be easily retorted to existing systems.

SQL injection has been used to extract customer and order information from e-commerce
databases, or bypass security mechanisms. The intuition behind such attacks is that
predefined logical expressions within a predefined query can be altered simply by injecting
operations that always result in true or false statements.

To prevent this from happening, two technologies have been introduced: one is improving the
programming techniques, another one is use the \api{PREPARE}d statement features supported
by many databases, which allows a client to pre-issue a template SQL query at the
beginning of a session. But these two approaches do not work very well. Then a concept
of instruction-set randomization is being introduced. For safeguarding systems against
any type of code-injection attack, by creating process-specific randomized instruction
sets (e.g., machine instructions) of the system executing potentially vulnerable software.
An attacker that does not know the key to the randomization algorithm will inject code
that is invalid for that randomized processor (and process), causing a runtime exception.
The same technique to the problem of SQL injection attacks. It creates randomized
instances of the SQL query language, by randomizing the template query inside the CGI
script and the database parser. To allow for easy retorting of our solution to existing
systems, it introduces a de-randomizing proxy, which converts randomized queries to
proper SQL queries for the database.
When this is the outcome, then standard keywords
lose their significance, and attacks are frustrated before they can even commence.

\subsubsection{SQLrand System Architecture}

Many web applications requiring users input, feed it into a pre-defined query.
After that, the query is handed to the database for execution. It is easy to make a mistake
when users input. Database system does not check the validation of the input data; hence,
it usually results in alteration to the database structure, corruption of data or revelation of
private and confidential information may be granted inadvertently.

\begin{figure}
	\centering
	\includegraphics[width=.7\textwidth]{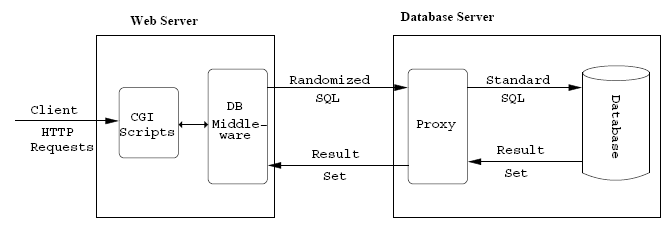}
	\caption{SQLrand System Architecture}
	\label{fig:sql-randomization-proxy}
\end{figure}

Here is one example:

\begin{verbatim}
select * from userTable where username=".$uid." and password = password (`. $pwd . ');
\end{verbatim}

Here the string variables \texttt{\$uid} and \texttt{\$pwd} come from the input from the user. The attacker
can set the \texttt{\$uid} variable to the string: \texttt{` or 1=1;--'}, it causes the SQL query to become:

\begin{verbatim}
select * from  userTable where username = "" or 1=1; -- and password = password(_any_text_);
\end{verbatim}

Notice here the double hyphen \texttt{--} comments out the remainder of the SQL query. It
means when there is a record in the \texttt{userTable} the query turns to be true, and bypasses the
password verification.
In order to avoid this happen, the application of Instruction-Set Randomization to the
SQL language: the SQL standard keywords are manipulated by appending a random
integer to them, one that an attacker cannot easily guess. Therefore, any malicious user
attempting an SQL injection attack would be thwarted, for the user input inserted into the
randomized query would always be classified as a set of non-keywords, resulting in an
invalid expression.
A difficult approach would be to modify the databases interpreter to accept the new set
of keywords.  Hence, there is the design consists of a proxy that sits between the client
and database server (see \xf{fig:sql-randomization-proxy}). The proxy may be on a separate machine. The proxy's
primary obligation is to decipher the random SQL query and then forward the SQL
command with the standard set of keywords to the database for computation. Another
benefit of the proxy is the concealment of database errors which may unveil the random
SQL keyword extension to the user.
A typical attack consists of a simple injection of SQL, hoping that the error message will
disclose a subset of the query or table information, which may be used to deduce
intuitively hidden properties of the database. By stripping away the randomization tags in
the proxy, we need not worry about the DBMS inadvertently exposing such information
through error messages; the DBMS itself never sees the randomization tags. Thus, to
ensure the security of the scheme, we only need to ensure that no messages generated by
the proxy itself are ever sent to the DBMS or the front-end server. Given that the proxy
itself is fairly simple, it seems possible to secure it against attacks. In the event that the
proxy is compromised, the database remains safe, assuming that other security measures
are in place.

\subsubsection{Implementation}

For the implementation, the proof-of-concept proxy server that sits between the web server and SQL server, de-randomizes
requests received from the web server, and conveys the query to the server. If an SQL injection attack has occurred, the proxy's
parser will fail to recognize the randomized query and will reject it. The randomized SQL parser utilized two popular tools for
writing compilers and parsers: \tool{flex} and \tool{yacc}. Capturing the encoded tokens required regular expressions that
matched each SQL keyword followed by zero or more digits. If properly encoded, the lexical analyzer strips the tokens extension
and returns it to the grammar for reassembly with the rest of the query. Otherwise, the token remains unaltered and is labeled
as an identifier. By default, \tool{flex} reads a source file, but the design required an array of characters as input. The
\api{YY\_INPUT} macro was re-defined to retrieve tokens from a character string introduced by the proxy. In the parsing phase, an error will be triggered. Either the developers SQL templates are incorrect or the users input unexpected data. If this happen, the parser returns \texttt{NULL}; otherwise, the de-randomized SQL string is returned. The parser was designed as a {\C} library.
With the parser completed, the communication protocol had to be established between the proxy and a database.

\subsubsection{Static, Dynamic Randomization Proxies}

\paragraph*{Static Randomization}
Here is an example, in the PHPBB v2.0.5, presented an opportunity to inject SQL into \file{viewtopic.php}. The proxy hardcodes the randomization key, which means the proxy, keeps the same randomization key every time for every query. For sometime, it was observed that the application displays an SQL query to the user by default when zero records are returned. Since an exception does not return any rows, the proxy's encoding key was revealed.

\paragraph*{Dynamic Randomization}

The programmer can develop a key randomization generator, which generates the different randomization keys, and implement the randomization key to the JDBC/ODBC driver side and proxy server side at the same time. If the randomization key changes, the keys on both sides also change. Each time the new keys is generated, the proxy receives a new query script from CGI script. If the proxy server met the SQL injection script, it returns zero records. Even the application displays the SQL query to the user (could be an attacker), the next time the randomized key changed from both sides of database middleware and the proxy server.  The previous randomized key is useless to the attacker. Compared to the static randomization, the dynamic randomization is, therefore, more secure.

\paragraph*{Key Management}

Here we present key management scheme to update keys
from time-to-time, with the idea borrowed from SNMPv3
key generation and update \cite{assi-inse7120-2007}.

\paragraph*{Key Generation}

We use the key generation idea of SNMPv3 to generate the randomization keys.
The localized keys are generated as follows.
We pick a password:

\begin{enumerate}
\item
User's password is expanded and hashed, producing \texttt{digest0},
which become's the user's key.
\item
\texttt{digest1 = Hash (digest0)}, where \texttt{digest1} is 16-octet (MD-5) or 20-octet (SHA-1).
      \texttt{authKey} for query/SQL keyword or in general authentication is \texttt{digest1}.
\item
Then, at each appliction deployment, a localized key is derived from \texttt{digest1}
by hashing \texttt{digest1} and some ID or random number, and becomes
so-called \texttt{digest2}.
We use a randomization number generator (see \xf{fig:localized-key}) to generate a random number, and take a hash of user key and random number to get the localized key \texttt{digest2}.
Application developer could append localized key \texttt{digest2} to the SQL script (schema or keywords).
\end{enumerate}

\paragraph*{Key Update}

To enhance security, keys are changed from time to time, because the key at the DB middleware side is not secure enough.
To replace an old key (\texttt{keyOld}) with a new key (\texttt{keyNew}) securily, the following steps are carried out \cite{assi-inse7120-2007}:

\begin{enumerate}
\item
Requestor generates a \texttt{random} number.

\item
Requestor computes a \texttt{digest = Hash(keyOld || random)}

\item
Requestor computes \texttt{delta = digest XOR keyNew}

\item
Requestor composes \texttt{protocolKeyChange = (random || delta)}

\item
Requestor sends message \texttt{setRequest(protocolKeyChange)}

\item
Receiver computes \texttt{digest = Hash(keyOld || random)}

\item
Receiver computes \texttt{keyNew = digest XOR delta}

\end{enumerate}

This works because \texttt{digest XOR delta = digest XOR (digest XOR keyNew) = keyNew}.
This is secure because the attacker presumably doesn't know \texttt{keyOld}, so such
an exchange is reasonably safe (in part also depends on the strength of the random
number generator).

\begin{figure}
	\centering
	\includegraphics[width=.7\textwidth]{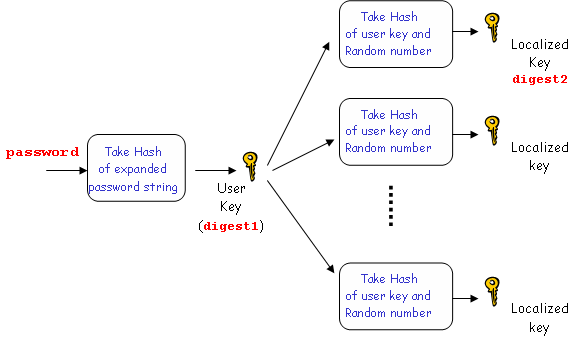}
	\caption{Localized Key Generation Solution}
	\label{fig:localized-key}
\end{figure}

The advantage of this method is application developer only maintains one key (one password).

\subsubsection{Parameters Summary}

\begin{enumerate}
\item
SQL script (query), Structure Query Language statement which been used to retrieve the information of database.
\item
Key (static or dynamic), the appended random number.
\item
User key, Take Hash of expanded password string.
\item
Localized key, hash of user key (\texttt{digest1}) and random number.
\end{enumerate}

%% file: methodology.tex
\chapter{Methodology}
\index{Methodology}

$Revision: 1.1.2.6 $

This section presents the framework design based on the studied methods, algorithms,
and techniques in this project.
The design methodology is primarily based on the algorithms and their parameters
presented earlier in Literature Review as well as a plug-in type of architecture
for various components whose implementation can be easily replaced. Thus the framework
presents a collection of interfaces for all technique types (CIA), followed by
their generic and concrete implementations. The concrete implementations came
from different open-source vendors and require adaptation to the framework, that's
why a layer of abstraction is introduced to adapt the data between algorithm
implementors and the framework. Further, to apply the framework to MARF and
HSQLDB, concrete security adapters are designed to make use of the JDSF ``injected''
into the core storage management components of both MARF (\api{marf.Storage.StorageManager})
and HSQLDB (\api{org.hsqldb.persist.Log}) where
they make sure the data hits the storage other than the main memory.

\section{Algorithms}

This section is a very brief, mostly pictographic summary of some
of the algorithms that were not presented earlier.

\subsection{Confidentiality}
\index{Methodology!confidentiality}

Basic CBC-DES encryption for confidentiality is in \xf{fig:cbc-crypt} and \xf{fig:cbc-decrypt}.

\begin{figure}
	\centering
	\includegraphics[width=.7\textwidth]{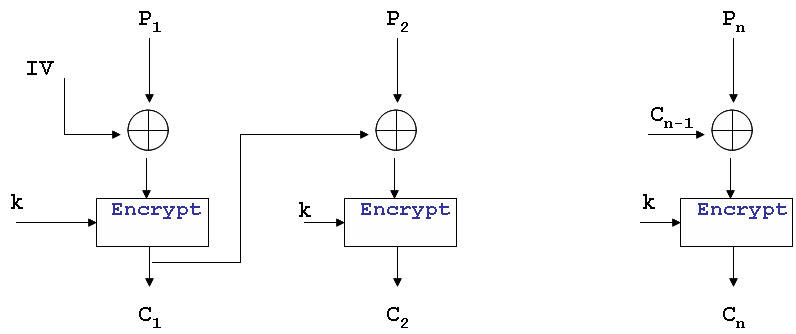}
	\caption{CBC Encryption of Data Elements \cite{assi-inse7120-2007}.}
	\label{fig:cbc-crypt}
\end{figure}

\begin{figure}
	\centering
	\includegraphics[width=.7\textwidth]{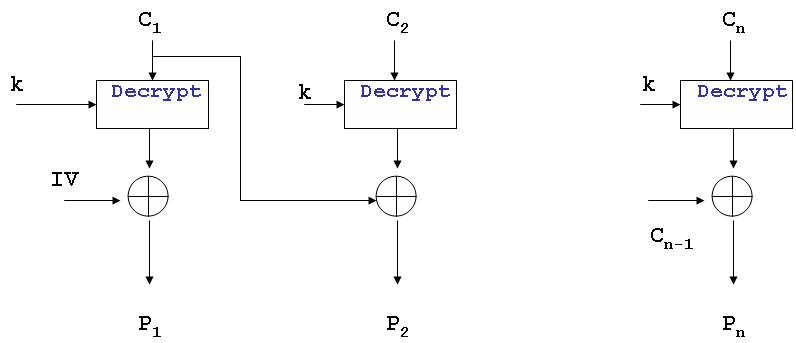}
	\caption{CBC Decryption of Data Elements \cite{assi-inse7120-2007}.}
	\label{fig:cbc-decrypt}
\end{figure}

\subsection{Integrity}
\index{Methodology!integrity}

For integrity-related algorithms please refer back to
\xs{sect:integrity} as well as information in \xf{fig:marf-security-integrity-integritylock}
for Integrity Lock modification, audio frequency masking in \xf{fig:marf-security-integrity-audiofrequencymasking},
watermark insertion (\xf{fig:marf-security-integrity-watermarkinsertion}) and
detection (\xf{fig:marf-security-integrity-watermarkdetection}) algorithms.

\subsection{Authentication}
\index{Methodology!authentication}

Basic HMAC-based authentication (and integrity) is in\xf{fig:hash-function} and in \xf{fig:hmac-auth-process}.

\begin{figure}
	\centering
	\includegraphics[width=.3\textwidth]{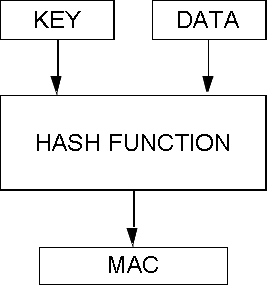}
	\caption{Basic HMAC-based Authentication \cite{assi-inse7120-2007}.}
	\label{fig:hash-function}
\end{figure}

\begin{figure}
	\centering
	\includegraphics[width=.7\textwidth]{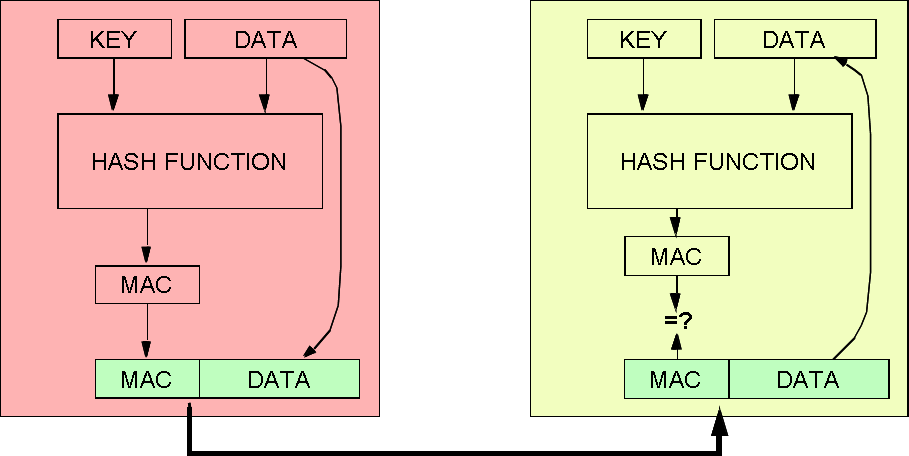}
	\caption{HMAC-based Authentication Process \cite{assi-inse7120-2007}.}
	\label{fig:hmac-auth-process}
\end{figure}

\subsection{SQL/Schema Randomization}
\index{Methodology!SQL/Schema randomization}

The SQL/schema randomization techniques of randomization proxies
and dynamic key updates were discussed in \xs{sect:sql-randomization}, implementation,
as well as the the pictographic algorithm is there in \xf{fig:sql-randomization-proxy}.

\section{Framework}
\index{Methodology!Framework Operation}

\subsection{General Operation}

In \xf{fig:auth-decr-data-store} is a general way the framework's particular adapters
(e.g. for MARF and HSQLDB) write the security-enhanced data based on the security
configuration options, set by the system administrator. The reading of the security-data
is the reverse process. A part of the operation on the diagram based on the SNMPv3
slides presented in \cite{assi-inse7120-2007}.

\begin{figure}
	\centering
	\includegraphics[width=\textwidth]{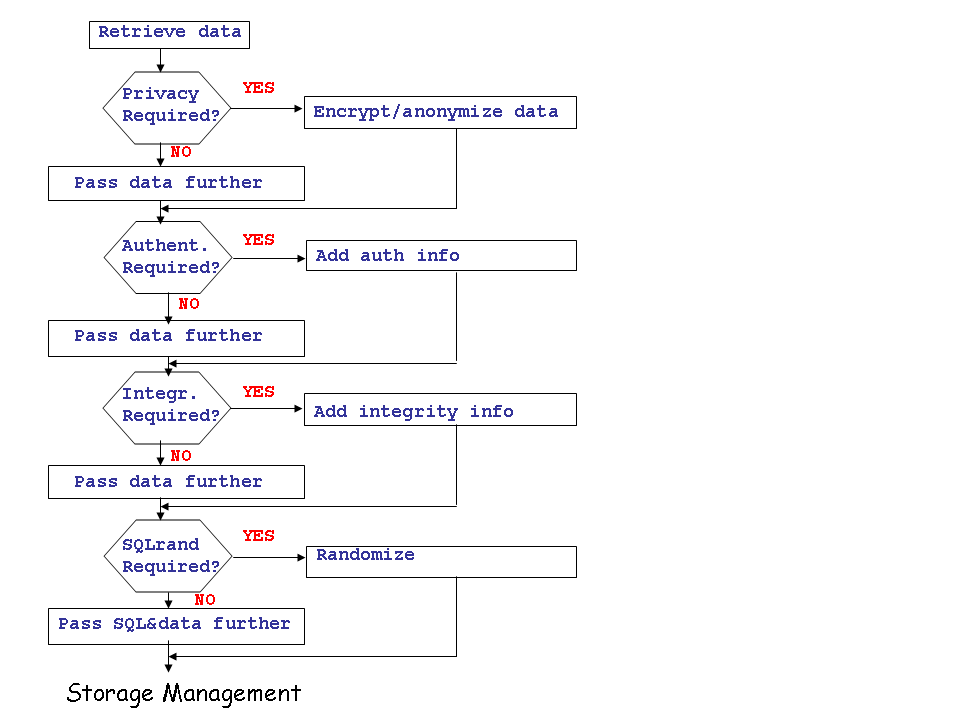}
	\caption{Writing Data With Security Options \cite{assi-inse7120-2007}.}
	\label{fig:auth-decr-data-store}
\end{figure}


\subsection{Design}

The typical MARF's packages (see \xf{fig:marf-dbsec-pkgs}) were extended with the two new packages
that constitute the JDSF: \api{database} and \api{marf.security}.
In \xf{fig:marf-security} are the primary packages and classes
that correspond to the studied aspects (CIA) and the utility, storage,
algorithms building blocks they rely upon. Since SQL/schema randomization is a
kind of a separate type of feature, it resides in its own package,
the contents of which is in \xf{fig:database-security-classes}.

The \api{marf.security.Configuration} class is populated from
the configuration file \file{security.properties}, and is set
by the system administrator. It represents the security options
desired by a given instance of the framework-enhanced data
management tool (e.g. MARF and HSQLDB).

The rest of the framework's backbone is captured by the main interfaces
and generic classes, followed by concrete implementation and stub
modules, and cryptographic algorithm providers. The interfaces allow
external to JDSF plug-ins, provided by external, third parties to be
able to extend and compare existing implementations if desired. The
interfaces are:

\begin{itemize}
\item \api{marf.security.confidentiality.IConfidentialityModule} in \xf{fig:marf-security-confidentiality}
\item \api{marf.security.integrity.IIntegrityModule} in \xf{fig:marf-security-integrity}
\item \api{marf.security.authentication.IAuthenticationModule} in \xf{fig:marf-security-authentication}
\item \api{marf.security.algorithms.IAlgorithmProvider} in \xf{fig:marf-security-algorithms}
\item \api{marf.security.Storage.ISecurityEnhancedObject} in \xf{fig:marf-security-storage}
\item \api{database.security.ISchemaRandomizationModule} in \xf{fig:database-security-classes}
\item \api{marf.security.adapters.ISecurityAdapter} in \xf{fig:marf-security-adapters}
\end{itemize}

In the \api{marf.security.algorithms} there are implementations of
well known cryptographic algorithms, such as CBC-DES, RSA, DSA, MD5,
and SHA1. The actual implementations in Java were provided by open-source
vendors, such as \cite{oos-java-cbc-des,oos-java-rsa,oos-java2s-dsa,oos-java-sun-dsa,oos-java-md5,oos-java-sha1}.
Since these implementations have sometimes little in common, integrating
it into the framework had to be abstracted by a common API of algorithm
providers (as in \xf{fig:marf-security-algorithms}, so the rest of the
framework does not depend on the vendors' API and can be replaced to
use another implementation easier when desired.

The most complexity goes into implementation and integration of the
framework into the actual data management tools, such as MARF and HSQLDB.
For this we provide their specific adapters (see \xf{fig:marf-security-adapters}): \api{marf.security.adapters.MARFSecurityAdapter}
that extends MARF-specific storage management and \api{marf.security.adapters.HSQLDBSecurityAdapter} likewise for HSQLDB, which are there to be
``injected'' into the original code wrapping storage management functions
of the original tools to mandatory go through the security-enhanced API.
The replaced and/or extended modules exactly are \api{marf.Storage.StorageManager} for MARF
and \api{org.hsqldb.persist.Log} for HSQLDB.

\begin{figure}
	\centering
	\includegraphics[width=\textwidth]{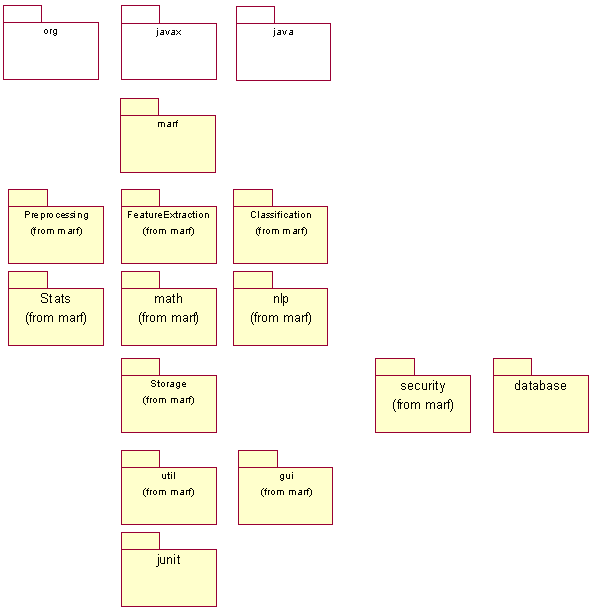}
	\caption{MARF Augmented with Security Database Packages.}
	\label{fig:marf-dbsec-pkgs}
\end{figure}

\begin{figure}
	\centering
	\includegraphics[width=.7\textwidth]{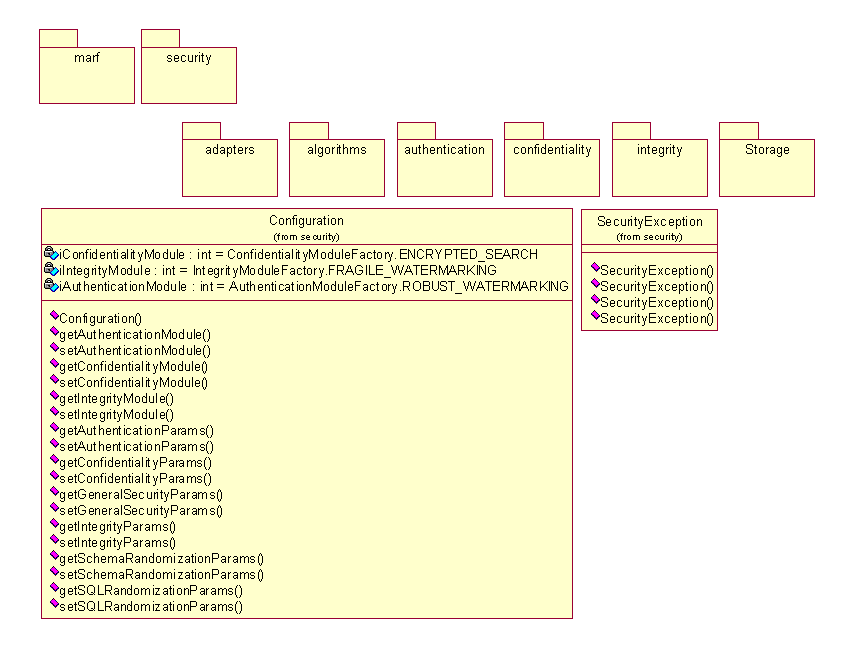}
	\caption{\api{marf.security} Package.}
	\label{fig:marf-security}
\end{figure}

\begin{figure}
	\centering
	\includegraphics[width=.7\textwidth]{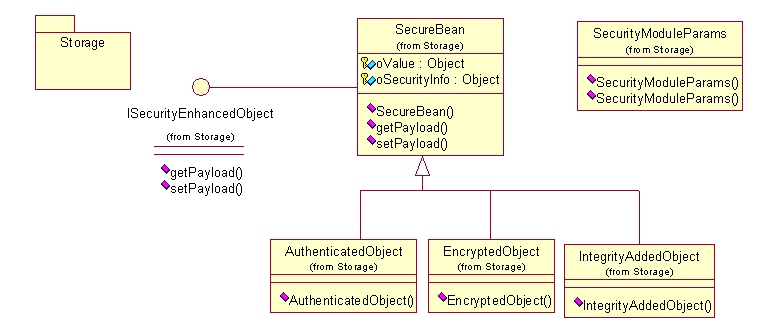}
	\caption{\api{marf.security.Storage} Package and Classes.}
	\label{fig:marf-security-storage}
\end{figure}

\begin{figure}
	\centering
	\includegraphics[width=.7\textwidth]{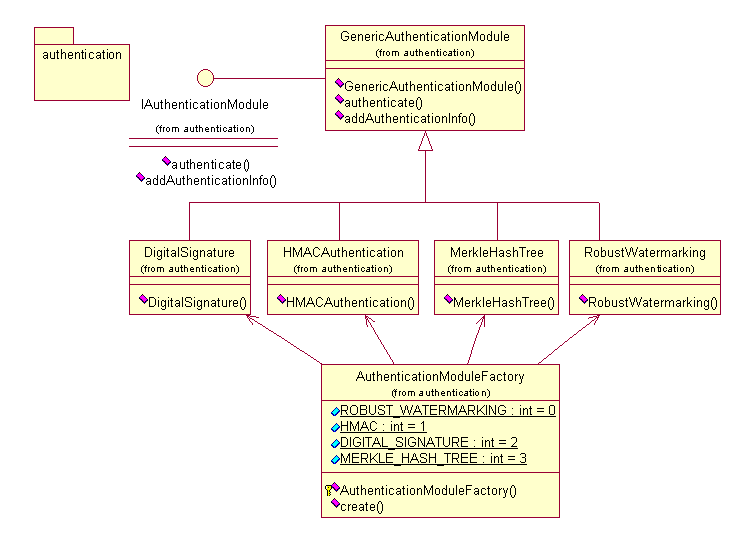}
	\caption{\api{marf.security.authentication} Package and Classes.}
	\label{fig:marf-security-authentication}
\end{figure}

\begin{figure}
	\centering
	\includegraphics[width=.7\textwidth]{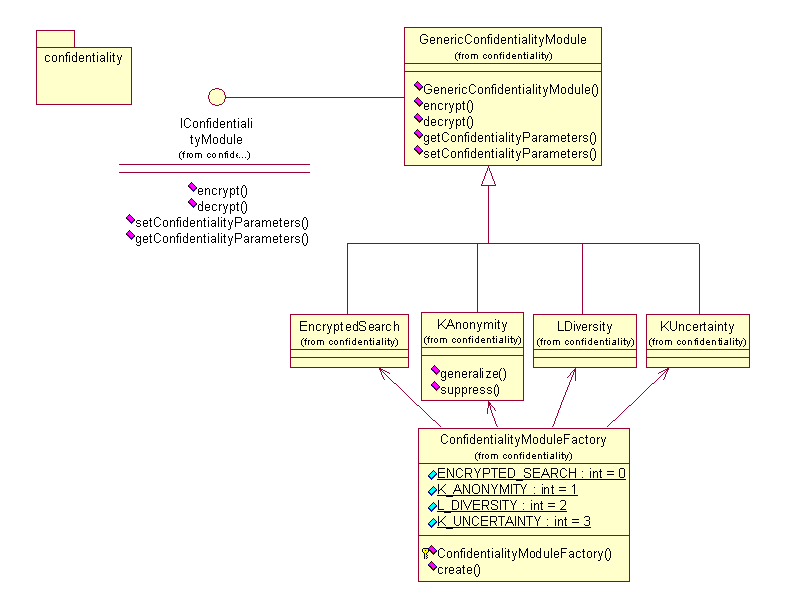}
	\caption{\api{marf.security.confidentiality} Package and Classes.}
	\label{fig:marf-security-confidentiality}
\end{figure}

\begin{figure}
	\centering
	\includegraphics[width=.7\textwidth]{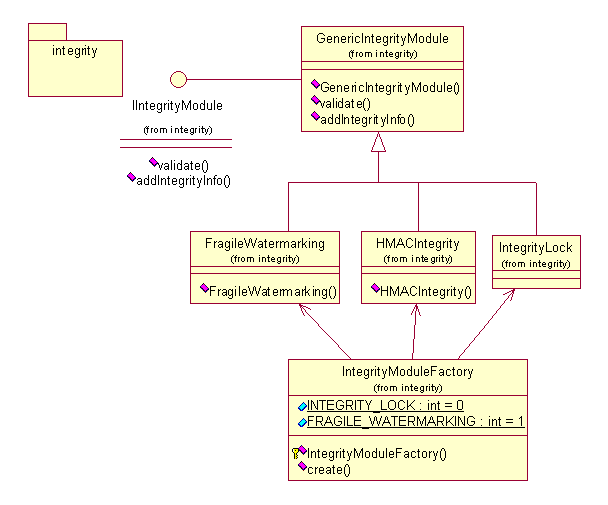}
	\caption{\api{marf.security.integrity} Package and Classes.}
	\label{fig:marf-security-integrity}
\end{figure}

\begin{figure}
	\centering
	\includegraphics[width=.7\textwidth]{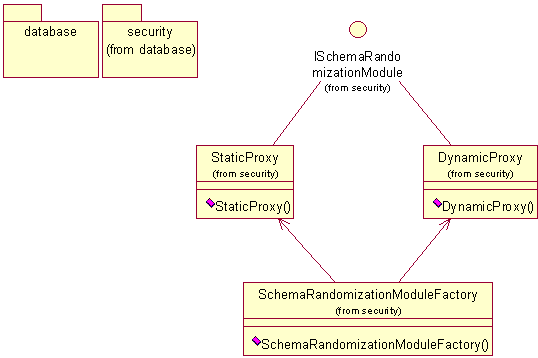}
	\caption{\api{database.security} Package and Classes.}
	\label{fig:database-security-classes}
\end{figure}

\begin{figure}
	\centering
	\includegraphics[width=.7\textwidth]{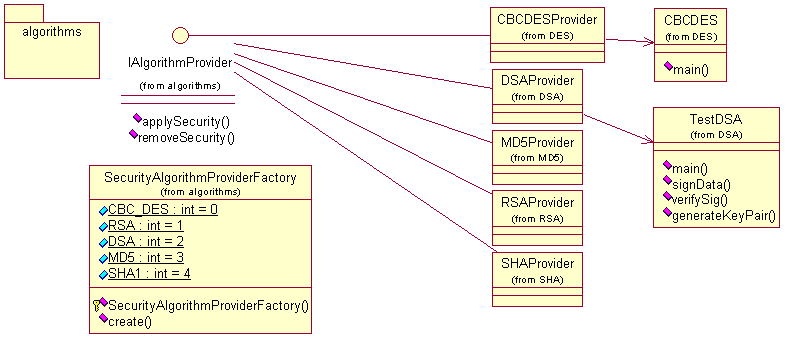}
	\caption{\api{marf.security.algorithms} Package and Classes.}
	\label{fig:marf-security-algorithms}
\end{figure}

\begin{figure}
	\centering
	\includegraphics[width=.7\textwidth]{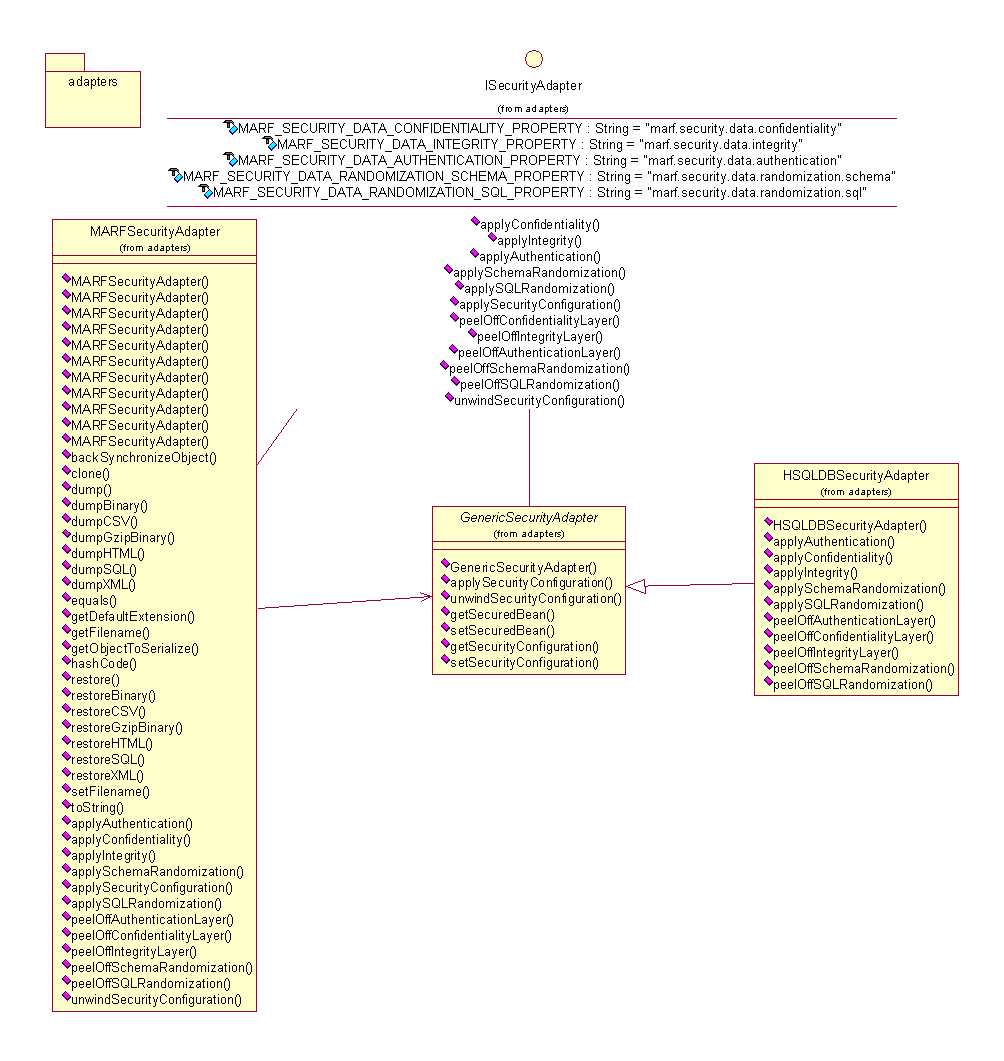}
	\caption{\api{marf.security.adapters} Package and Classes.}
	\label{fig:marf-security-adapters}
\end{figure}


%% file: conclusion.tex
\chapter{Conclusion}
\index{Conclusion}

$Revision: 1.1.2.9 $

\section{Overview}

Confidentiality aspect: we explored a few popular techniques for
encrypted search, $k$-anonymity, $l$-diversity, etc. We treat
non-encryption anonymization data as encryption in our framework,
i.e. we use the same API for both except in the latter case
``encryption'' means for example ``generalization'' or ``suppression''
(or even ``compression'').

Integrity aspect: in the integrity aspect, we discuss four types of digital watermarking techniques,
which are audio watermarking, image watermarking, video watermarking, and relational database watermarking.
Moreover, we explain their methodologies and algorithms in some detail.
The goal of our project is to construct a security environment for MARF and HSQLDB by 4 aspects.
Since MARF framework is primarily an audio recognition system, for integrity aspect, we can implement
audio watermarking technology in audio file database system and relational database watermarking to secure whole attributes, tuples and relations in database.
We create parameters of algorithms and design packages with classes for future implementation.

Authentication aspect:
we mention a way for authenticated indexing schemes for aggregation queries.
We provided a structure with reasonable query performance in static environments.
However, it has increased space utilization for sparse databases and high update overhead.
Therefore, we presented structures for dynamic settings that gracefully adapt to data updates and have better space utilization for sparse datasets. We also showed how to extend these techniques to handle multiple aggregates and multiple selection predicates per query. For future work, we plan to explore solutions for holistic aggregates and investigate the application of our techniques to authenticate data cubes in OLAP system.

SQL randomization aspect: we have presented two different ways of implementing randomization, static randomization proxy and dynamic randomization proxy, and extend to the HSQLDB JDBC/ODBC driver that also performs SQL randomization. This system is broadly applicable (web-based dynamic content systems are widely deployed), addresses a critical security problem (SQL injection attacks), implements an innovative idea (SQL randomization), and is an improvement over previous efforts. The design is easy to use, portable to other database middleware drivers, should have a small performance impact, and makes SQL injection attacks infeasible.

Framework: JDSF's operation was designed to allow addition of any number of algorithms or techniques to
add as plug-ins for comparative study or when better techniques become available.
The parameters and the configuration of the framework were made available from the survey/research study
of the database security techniques presented earlier.
It is also general enough to expand beyond MARF and HSQLDB, and as a result
the open source community can benefit as a whole.

\section{Open-Source}

JDSF, just like MARF and HSQLDB, is open source and is host at \url{SourceForge.net}
under umbrella of MARF, in its CVS repository. The latest CVS revision of it and
MARF is under the INSE691A branch in the MARF's repository. Non-developers can
access it as follows:
\begin{verbatim}
cvs -z3 -d:pserver:anonymous@marf.cvs.sf.net:/cvsroot/marf co -rINSE691A marf
\end{verbatim}
\noindent
or browse it all on-line at \url{http://marf.cvs.sf.net} under the same branch.

\section{Future Work}

As a future work we plan on continuing our open-source development
effort of the framework and fully integrating it into MARF and HSQLDB,
along with comprehensive testing. Additionally, we plan on publishing
our results and contributing the framework to be introduced into the
core of HSQLDB and MARF.

\section{Acknowledgments}

\begin{itemize}
\item Dr. Lingyu Wang
\item Database security researchers
\item Open-Source Community
\item Fellow classmates
\end{itemize}


%% file: references.tex
\addcontentsline{toc}{chapter}{Bibliography}

\bibliography{report}
\bibliographystyle{alpha}
